\documentclass{webofc}
\usepackage[varg]{txfonts}   
\usepackage{slashed}
\newcommand*\colvec[1]{\begin{pmatrix}#1\end{pmatrix}}
\begin{document}
\title{Chiral asymmetry during the EWPT from CP-violating scattering off bubble walls}
%
%

\author{\firstname{Alejandro} \lastname{Ayala}\inst{1,2}\fnsep\thanks{\email{ayala@nucleares.unam.mx}} \and
        \firstname{L. A.} \lastname{Hernández}\inst{1}\fnsep\thanks{\email{luis.hernandezr@correo.nucleares.unam.mx}} \and
        \firstname{Jordi} \lastname{Salinas}\inst{1}\fnsep\thanks{\email{jordissm@ciencias.unam.mx}}
}

\institute{Instituto de Ciencias Nucleares, Universidad Nacional Autónoma de México, Apartado Postal 70-543, Ciudad de México 04510, México
\and
Centre for Theoretical and Mathematical Physics, and Department of Physics, University of Cape Town, Rondebosch 7700, South Africa
        }

\abstract{%
We compute a net electric current during a first order EWPT arising from the asymmetric propagation of fermion chiral modes due to a CP-violating interaction with the Higgs. The interaction is quantified in terms of a CP-violating phase in the bubble wall that separate both phases. We comment on the possibility of this current to generate a seed magnetic field and its implications for primordial magnetogenesis in the early Universe.
}
\maketitle
\section{Introduction}
\label{intro}
The physics of the Early Universe remains unknown on many of its aspects. One of the long standing problems of this era is the generation of the matter-antimatter asymmetry, the so called, baryogenesis. Observations from $\gamma$-ray telescopes indicate that there are no sources of this annihilation processes happening in the visible Universe. This situation leads to think that an asymmetry was created soon after the Big Bang, leaving little room for antimatter to spread. \par
Another outstanding problem in the realm of cosmology is the puzzling presence of large-scale magnetic fields, coherent up to galactic and intergalactic distances. Recent evidence has allowed to observe magnetic fields coherent up to galactic distances with a strength of a few $\mu$G, one millionth part of Earth's magnetic field. Despite their ample presence in the Universe, the origin of the magnetic fields of such astronomical scales remains a mystery \cite{Durrer}. \par
The observational data suggest that the fields that are seen are a product of the amplification of seed magnetic fields through some dynamo process \cite{Zucca}. Furthermore, the experimental limits obtained from high redshift galaxies seem to indicate that this dynamo would have little time to act and would require field strengths of $10^{-6}$G to $10^{-30}$G, depending on the model \cite{Kahniashvili}.\par
Several scenarios have been studied to explain the possible mechanisms that may have given rise to these seed fields. The two common categories that are used to classify the possible methods of generating the seed fields are astrophysical sources which tend to amplify magnetic fields from stars and other astrophysical objects up to galactic lengths, and cosmological or primordial sources that propose distinct forms of generating large-scale magnetic fields before the galaxies are themselves formed, thus explaining at the same time the intracluster and void magnetic fields. One of the plausible cases of primordial generation of magnetic fields takes place during the electroweak phase transition (EWPT) when the Universe had temperatures of the order of 100 GeV. It is known that the EWPT is weakly first order as dictated by the Higgs mass, and can become stronger in the presence of magnetic fields \cite{Carrington, Subramanian}. 
\section{Dirac equation with a complex mass}\label{sec: equation}
The bubble nucleation during the EWPT separates the vacuum phases and the membrane separating the inside and outside of the bubble is called the wall. The properties of the wall depend on those of the electroweak effective potential. Taking into account the need for CP violation and the departure from thermal equilibrium from Sakharov's conditions, we set a complex mass that introduces explicit CP symmetry violation during the EWPT \cite{Ayala}. We cast the problem of fermion scattering off the bubble wall in terms of solving a Dirac equation with a position dependent fermion mass $m(z)$, proportional to the Higgs field. Explicitly,
\begin{equation}
	m(z)=m_0 e^{i\phi}\Theta(z),
\end{equation}
where $\phi$ is a phase, $\Theta(z)$ is the step function and $z$ is the axis perpendicular to the bubble wall. As CP-violation implies that left- and right-handed chirality modes of a fermion spinor $\psi$ couple to the Higgs field with $m(z)$ and $m^*(z)$, respectively, the Dirac equation is written as
\begin{equation}\label{DiracEq}
\left\{ i\slashed{\partial}-m^*\frac{1}{2}(1-\gamma^5)-m\frac{1}{2}(1+\gamma^5)\right\}\Psi =0.
\end{equation}

\section{Solution of the Dirac equation}\label{sec: solving}
Each of the fermion spinor components in Eq. (\ref{DiracEq}) have to satisfy the Klein-Gordon equation. Therefore, a general solution is of the form
\begin{equation}\label{Ansatz}
\left\{ i\slashed{\partial}+m^*\frac{1}{2}(1-\gamma^5)+m\frac{1}{2}(1+\gamma^5)\right\}\mathit{\Phi} =0,
\end{equation}
and when Eq. (\ref{Ansatz}) is substituted into Eq. (\ref{DiracEq}), we obtain the corresponding KG-like equations
\begin{equation}\label{KG}
\left\{ -\slashed{\partial}^2-i\gamma^3m_0e^{-i\phi}\delta(z)\frac{1}{2}(1-\gamma^5) -i\gamma^3m_0e^{i\phi}\delta(z)\frac{1}{2}(1+\gamma^5)-m_0^2\Theta(z)\right\}\mathit{\Phi} =0.
\end{equation}
By writing $\mathit{\Phi}(\vec{x},t)$ as $\mathit{\Phi}(\vec{x},t)=\xi(x,y)e^{-iEt}\Phi(z)$ and focusing on the solution for the $z$-direction we get
\begin{equation}\label{EqZ}
\left\{ E^2-m_0^2\Theta(z)+d_z^2 -im_0\delta(z)\gamma^3\left[ e^{-i\phi}\frac{1}{2}(1-\gamma^5)+e^{i\phi}\frac{1}{2}(1+\gamma^5)\right] \right\}\Phi(z) =0.
\end{equation}
Expanding the function $\Phi(z)$ into a linear combination of the eigenspinors of $\gamma^3$ which are
\[u^1_{\pm}=\colvec{1\\0\\\pm i\\0}, \qquad u^2_{\pm}=\colvec{0\\1\\0\\\mp i}\]
where $\{1,2\}$ correspond to the spin projection and $\{+,-\}$ to the positive and negative energies, respectively. Using these spinors, $\Phi(z)$ can be written as
\begin{equation}
\Phi(z)=\Phi^1_+(z)u_+^1+\Phi^1_-(z)u_-^1+\Phi^2_+(z)u_+^2+\Phi^2_-(z)u_-^2.
\end{equation}
By substituting this into Eq. (\ref{EqZ}), we obtain the system of coupled equations
\begin{align}
&[E^2-m_0^2\Theta(z)+d_z^2+m_0\delta(z)\cos\phi]\mathit{\Phi}^1_+(z)=-im_0\delta\sin\phi\mathit{\Phi}^1_-(z)\\
&[E^2-m_0^2\Theta(z)+d_z^2-m_0\delta(z)\cos\phi]\mathit{\Phi}^1_-(z)=im_0\delta\sin\phi\mathit{\Phi}^1_+(z)
\end{align}
with a similar set for $\mathit{\Phi}^2_+$ and $\mathit{\Phi}^2_-$. We note that when the fermion is in each of the phases, the equations decouple. 
The solution of the system of decoupled equations is given in terms of plane waves. We physically 
foresee that coefficients $G_{\pm}$ that represent an incoming wave from the RHS to the delta potential must be null. Solving the system of equations for the coefficients and implementing the continuity conditions, we get the following solutions.\\
For $z<0$,
\begin{align}
\mathit{\Phi}_+^{1,2}&=A_+e^{iEz}+\left[\left(\frac{im_0\cos\phi}{E+\sqrt{E^2-m_0^2}}\right)A_+-\left(\frac{m_0\sin\phi}{E+\sqrt{E^2-m_0^2}}\right)A_-\right]e^{-iEz}, \nonumber\\
\mathit{\Phi}_-^{1,2}&=A_-e^{iEz}+\left[\left(\frac{m_0\sin\phi}{E+\sqrt{E^2-m_0^2}}\right)A_+-\left(\frac{im_0\cos\phi}{E+\sqrt{E^2-m_0^2}}\right)A_-\right]e^{-iEz},\label{Solz<0}
\end{align}
For $z>0$,
\begin{align}
\mathit{\Phi}_+^{1,2}&=\left[\left(1+\frac{im_0\cos\phi}{E+\sqrt{E^2-m_0^2}}\right)A_+-\left(\frac{m_0\sin\phi}{E+\sqrt{E^2-m_0^2}}\right)A_-\right]e^{i\sqrt{E^2-m_0^2}z}, \nonumber\\
\mathit{\Phi}_-^{1,2}&=\left[\left(\frac{m_0\sin\phi}{E+\sqrt{E^2-m_0^2}}\right)A_++\left(1-\frac{im_0\cos\phi}{E+\sqrt{E^2-m_0^2}}\right)A_-\right]e^{i\sqrt{E^2-m_0^2}z}.
\end{align}
\section{Transmission coefficients}\label{sec: coefficients}
\begin{figure}[h]
\centering
\begin{minipage}{0.45\textwidth}
	\centering
	\includegraphics[width=5.5cm,clip]{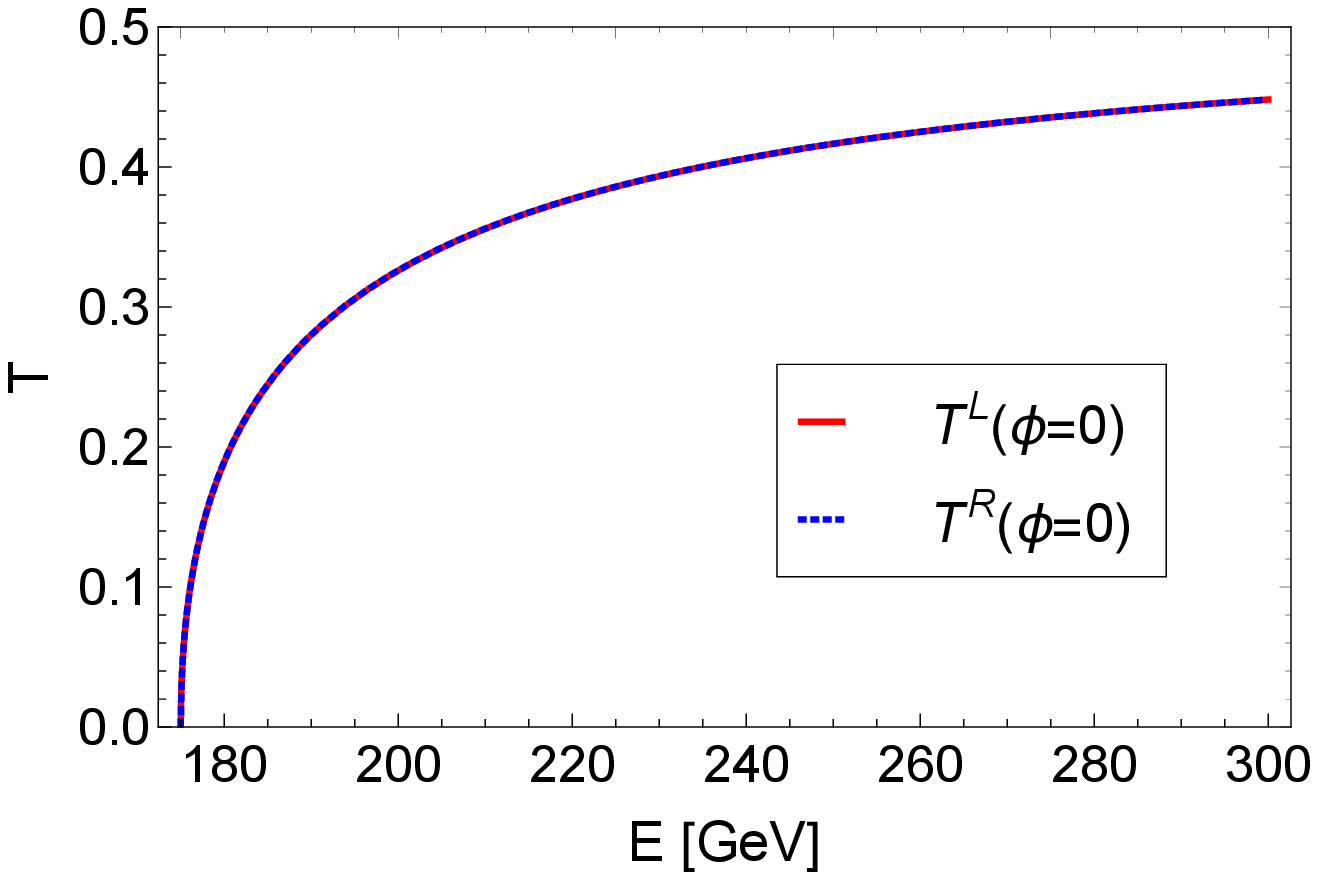}
	\caption{Chiral transmission coefficients for the case where $\phi=0$. As expected, there is no asymmetry present between both quantities.}
	\label{fig1}      
\end{minipage}\hspace{2mm}
\begin{minipage}{0.45\textwidth}
	\centering
	\includegraphics[width=5.5cm,clip]{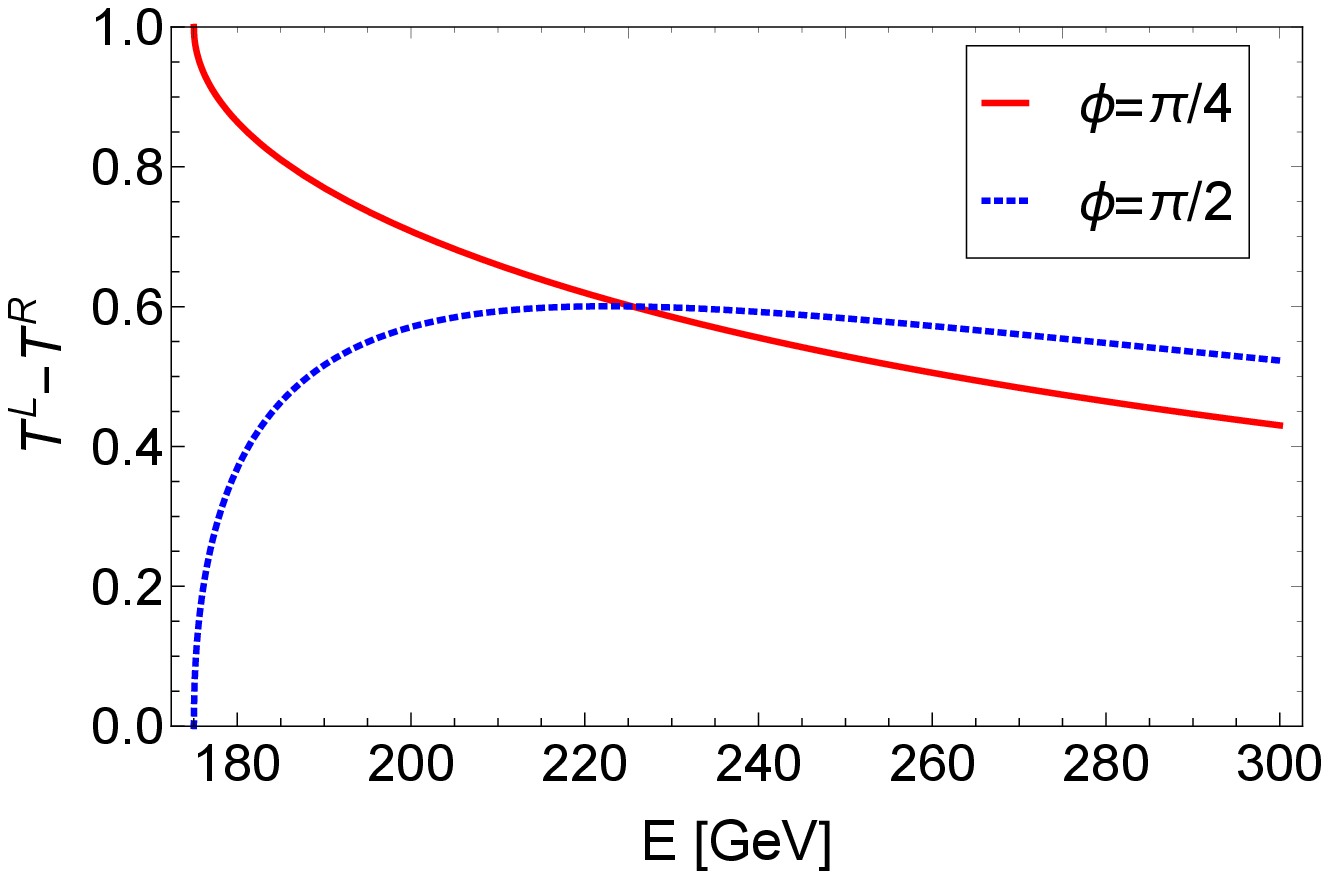}
	\caption{Chiral transmission coefficients difference for $\phi=\pi/4$ and $\phi=\pi/2$. The asymmetry is energy- and phase-dependent.}
	\label{fig2}       
\end{minipage}

\end{figure}

The fermion spinor $\psi(z)$ can be decomposed into its incident, reflected and transmitted components from which we can extract the corresponding currents. With these currents, it is also possible to calculate the transmission and reflection coefficients defined as per $R=J_{\text{ref}}/J_{\text{inc}}$ and $T=J_{\text{tra}}/J_{\text{inc}}$. Moreover, by applying the chiral projectors to the spinor, we can specifically obtain the distinct chirality transmission coefficients $T^L$ and $T^R$ which are found to be
\begin{align}
&T^L=\frac{1}{E\left( E+\sqrt{E^2-m_0^2}\right)^2}\cdot\Bigg[+m_0^3\sin\phi\cos\phi   + \sqrt{E^2-m_0^2}\left( E+\sqrt{E^2-m_0^2}\right)\left(E+m_0 \sin\phi\right)\Bigg] \nonumber \\
&T^R=\frac{1}{E\left( E+\sqrt{E^2-m_0^2}\right)^2}\cdot\Bigg[ -m_0^3\sin\phi\cos\phi + \sqrt{E^2-m_0^2}\left( E+\sqrt{E^2-m_0^2}\right)\left(E-m_0 \sin\phi\right) \Bigg]. \label{eq:coefficients}
\end{align}
Equation (\ref{eq:coefficients}) contains the information on the ratio of the number of fermions for each chiral mode being transmitted through the wall to the incident ones. The asymmetry between the modes is quantified in terms of the phase $\phi$. Figure \ref{fig1} displays the chiral transmission coefficients for the trivial case $\phi=0$ as a function of energy, where both probabilities coincide, as expected. Figure \ref{fig2} shows the left- and right- handed transmission coefficient difference for two distinct values of $\phi$, demonstrating the asymmetry between both quantities. 
\section{Summary and conclusions}\label{sec: summary}
In the thin-wall approximation, where the fermion mean free path
is far greater than the bubble wall thickness, fermions
behave as particles interacting only with the walls of a bubble
nucleated during the EWPT. We have showed that the introduction of a CP-violating interaction with the bubble through a complex phase in the fermion mass can generate a net injection of chiral flux that can be quantified in terms of the difference between chiral transmission coefficients. The above implies that for fermions with an electric charge $q$, a local electric current transverse to the wall is generated. \par 
In the case considered in this work, electric currents are created normal to the bubble walls with varying intensity, dependent to the particle's energy. These inhomogeneities can induce a helical magnetic field, a property that is crucial for possible future amplification processes \cite{Brandemburg}. The asymmetry is energy dependent and maximal for $\phi \simeq \pi/2$ and can be thought of to come from the complex phase in the CKM matrix. Although the magnitude of such a phase is consistent with current observations \cite{PDG}, one can think that the asymmetry might not be maximal.

%
%
%

\end{document}